\documentclass[journal]{IEEEtranTIE_authors_version}

\usepackage{MCLabPaper_Minimal}
\usepackage{rotating}
\usepackage{mdframed}
\usepackage{booktabs}

\begin{document}


\newcommand{\ArticleDisclaimer}{%
	This article appears in IEEE Transactions on Industrial Electronics, 2021. 
	DOI: \href{http://doi.org/10.1109/TIE.2021.3102431}{10.1109/TIE.2021.3102431}
}

\author{Igor~Melatti, Federico~Mari, Toni~Mancini, Milan~Prodanovic,~\IEEEmembership{Member,~IEEE}, and Enrico Tronci.
\thanks{Authors Melatti, Mancini and Tronci are with the Sapienza University of Rome, via Salaria 113, 00198 Rome, Italy. Author Mari is with the University of Rome ``Foro Italico'', Viale del Foro Italico, 00135 Rome, Italy. Author Prodanovic is with Electrical Systems Unit of IMDEA Energy Institute, Avda. Ram\'{o}n de la Sagra, M\'{o}stoles Technology Park, Madrid 28935, Spain. This work was partially supported by: 
Research  programme  S2018/EMT-4366 PROMINT-CAM  from  Madrid  Government;
Italian Ministry of University and Research under grant ``Dipartimenti di eccellenza 2018--2022'' of the Department of Computer Science of Sapienza University of Rome; 
EC FP7 project SmartHG; 
INdAM ``GNCS Project 2019''.

\bigskip

\begin{mdframed}
\begin{center}
	\ArticleDisclaimer
\end{center}
\end{mdframed}
}}


\title{A Two-Layer Near-Optimal Strategy for~Substation Constraint Management via~Home~Batteries}

\markboth{\ArticleDisclaimer}%
{Melatti \MakeLowercase{\textit{et al.}}: }

\maketitle



  \NewAcronym{AMI}{}{}{Advanced Metering Infrastructure}{}
  \NewAcronym{AMD}{}{}{Average of Residential Users Missed EV Deadlines Fractions}{}
  \NewAcronym{AMPC}{}{}{Adaptive Model Predictive Control}{}
  \NewAcronym{AMST}{AvgSolTime}{}{Average MILP Solving Time}{}
  \NewAcronym{ARS}{}{}{Average of Residential Users Saving}{}
  \NewAcronym{BEM}{}{}{Battery Energy Manager}{}
  \NewAcronym{BEP}{}{}{Break Even Point}{}
  \NewAcronym{BMA}{BattAccuracy}{}{Battery Model Accuracy}{}
  \NewAcronym{DADOBR}{DemOutRedOpt}{}{Optimal (Centralised) Aggregated Demand Outside Bounds Reduction w.r.t. Historical Demand}{}
  \NewAcronym{DALB}{DANCA}{}{Demand-Aware Network Constraint mAnager}{}
  \NewAcronym{SDALB}{SDANCA}{}{Simple Demand-Aware Network Constraint mAnager}{}
  \NewAcronym{DAPP}{ADAPT}{}{DemAnD--Aware Power limiT}{}
  \NewAcronym{DER}{}{}{Distributed Energy Resources}{}
  \NewAcronym{DG}{}{}{Distributed Generation}{}
  \NewAcronym{DLC}{}{}{Direct Load Control}{}
  \NewAcronym{DR}{ADR}{}{Autonomous Demand Response}{}
  \NewAcronym{DSM}{}{}{Demand Side Management}{}
  \NewAcronym{DSO}{}{}{Distribution System Operator}{}
  \NewAcronym{DSRA}{}{}{Demand Side Response Aggregator}{}
  \NewAcronym{EBR}{LAHEMS}{}{Lightweight Adaptive Home Energy Management System}{}
  \NewAcronym{EDN}{}{}{Electrical Distribution Network}{}
  \NewAcronym{EVT}{}{}{EDN Virtual Tomography}{}
  \NewAcronym{ESS}{}{}{Energy Storage System}{}
 \NewAcronym{GLPK}{}{}{GNU Linear Programming Kit}{}
  \NewAcronym{HEMS}{}{}{Home Energy Management System}{}
  \NewAcronym{HDQ}{}{}{High-Speed Data Queue}{}
  \NewAcronym{HVAC}{}{}{Heating, Ventilation and Air Conditioning}{}
  \NewAcronym{IBR}{}{}{Inclining Block Rate}{}
  \NewAcronym{KPI}{}{}{Key Performance Indicator}{}
  \NewAcronym{LADOBR}{DemOutRed}{}{DANCA (Hierarchical) Aggregated Demand Outside Bounds Reduction w.r.t. Historical Demand}{}
  \NewAcronym{LP}{}{}{Linear Programming}{}
  \NewAcronym{MMD}{MissDeadl}{}{Missed Deadline for MILP Solving}{}
  \NewAcronym{MILP}{}{}{Mixed Integer Linear Programming}{}
  \NewAcronym{MPC}{}{}{Model Predictive Control}{}
  \NewAcronym{MPDF}{UserDiscomfort}{}{Missed EV Deadlines}{}
  \NewAcronym{MST}{SolvingTime}{}{MILP Solving Time}{}
  \NewAcronym{MMST}{MaxSolvingTime}{}{Maximum MILP Solving Time}{}
  \NewAcronym{NHC}{HorChange}{}{Horizon Changes}{}
  \NewAcronym{OMPDF}{RetailDiscomfort}{}{Overall Missed EV Deadlines Fraction}{}
  \NewAcronym{ONB}{Dimensioning}{}{Optimal Number of Batteries}{}
  \NewAcronym{PPSV}{}{}{Price Policy Safety Verification}{}
  \NewAcronym{PEV}{EV}{}{
  Electric Vehicle}{}
  \NewAcronym{PHIL}{}{}{Power Hardware-In-the-Loop}{}
  \NewAcronym{PMA}{PEVAccuracy}{}{EV Model Accuracy}{}
  \NewAcronym{PVP}{}{}{Photovoltaic Panel}{}
  \NewAcronym{RTP}{}{}{Real-Time Price}{}
  \NewAcronym{RUSF}{UserNetSaving}{}{Residential Users Net Saving Percentage}{}
  \NewAcronym{SE}{}{}{State Estimation}{}
  \NewAcronym{SEIL}{}{}{Smart Energy Integration Lab}{}
  \NewAcronym{SCADA}{}{}{Supervisory Control And Data Acquisition}  {}
  \NewAcronym{SCM}{}{}{Substation Constraints Management}  {}
  \NewAcronym{SOC}{SoC}{}{State of Charge}  {}
  \NewAcronym{TD}{T\&D}{}{Trasmission \& Distribution}{}
  \NewAcronym{ToU}{}{}{Time of Use}{}
  \NewAcronym{UART}{}{}{Universal Asynchronous Receiver/Transmitter}{}
  \NewAcronym{UDP}{}{}{Usage-based Dynamic Pricing}{}
  \NewAcronym{USF}{RetailSaving}{}{Energy Retailer Saving Fraction}{}
  \NewAcronym{V2G}{}{}{Vehicle to Grid}{}
  \NewAcronym{V2H}{}{}{Vehicle to Home}{}
  \NewAcronym{WLS}{}{}{Weighted Least Squares}{}

\begin{abstract}

  Within electrical distribution networks, substation constraints management requires that aggregated power demand from residential users is kept within suitable bounds. Efficiency of substation constraints management can be measured as the reduction of constraints violations w.r.t. unmanaged demand. Home batteries hold the promise of enabling efficient and user-oblivious substation constraints management. Centralized control of home batteries would achieve optimal efficiency. However, it is hardly acceptable by users, since service providers (e.g., utilities or aggregators) would directly control batteries at user premises. Unfortunately, devising efficient hierarchical control strategies, thus overcoming the above problem, is far from easy.
We present a novel two-layer control strategy for home batteries that avoids direct control of home devices by the service provider and at the same time yields near-optimal substation constraints management efficiency. Our simulation results on field data from 62 households in Denmark show that the substation constraints management efficiency achieved with our approach is at least 82\% of the one obtained with a theoretical optimal centralized strategy.

\end{abstract}

\section*{Nomenclature}

\begin{description}

  \item[$s$] \Acs{EDN} substation

  \item[$T_s$] set of time slots for substation $s$ 

  \item[$U$] set of houses (users)

  \item[$t$] time slot (element of $T$)

  \item[$u$] house (element of $U$)

  \item[$T_u$] set of time slots for house $u$; note that, for all notation depending on an house index $u$, if the house is understood, $u$ is not shown

  \item[$T_{u, P}$] set of time slots in $T_u$ in which the \Acs{PEV} is plugged-in on house $u$ 

  \item[$\tau_l$] duration (in minutes) of time slots in $T_s$

  \item[$\tau_s$] time (in minutes) between two optimisation decisions on substation

  \item[$H_s = \frac{\tau_s}{\tau_l}$] optimisation horizon for substation

  \item[$\tau$] time (in minutes) between two optimisation decisions on houses 

  \item[$Q_{u, E}, m_{u, E}, M_{u, E}, \alpha_E, \beta_E$] maximum capacity (in kWh), minimum and maximum power rate (in kW), charge and discharge efficiency of \Acs{ESS} in  $u$ 

  \item[$Q_{u, P}, m_{u, P}, M_{u, P}, \alpha_P, \beta_P$] maximum capacity (in kWh), minimum and maximum power rate (in kW), charge and discharge efficiency of battery \Acs{PEV} in  $u$ 

  \item[$C_u^{low}, C_u^{high}$] minimum and maximum power demand (in kW) from energy contract of  $u$

  \item[$H_i, H, H_{\delta}$] initial optimisation horizon, current optimization horizon and optimization horizon changing step for computation in houses

  \item[$\zeta$] deadline (in minutes) to complete a single optimisation computation in houses

  \item[$P_{s}^{low}(t), P_{s}^{high}(t)$] lower and upper desired power bounds (in kW) for substation $s$ in  $t$

  \item[$\tilde{Q}_{u, E}, \tilde{Q}_{u, P}$] current \Acs{SOC} (in kWh) of \Acs{ESS} and \Acs{PEV} 
    in  $u$

  \item[${\cal D}_{u, P}$] deadline (in minutes) for \Acs{PEV} complete recharge in  $u$

  \item[$d_{u}(t)$] forecasted power demand (in kW) of  $u$ in  $t$

  \item[$b_{u, E}(t)$] \Acs{SOC} (in kWh) of \Acs{ESS} in  $u$ at  $t$

  \item[$a_{u, E}(t)$] charge or discharge action (in kW) for the \Acs{ESS} in  $u$ at  $t$

  \item[$P_{u}^{low}(t), P_{u}^{high}(t)$] lower and upper power limits (in kW) for  $u$ in  $t$

  \item[$\Delta^{low}(t), \Delta^{high}(t)$] aggregated power (in kW) which exceeds substation  
    lower and upper bounds 
    in  $t$

  \item[$e_u(t)$] resulting power demand (in kW) in  $u$ at  $t$

  \item[$a^{\rm ch}_{u, E}(t), a^{\rm dis}_{u, E}(t)$] charge and discharge actions (in kW) for the \Acs{ESS} in  $u$ at  $t$

  \item[$a^{\rm ch}_{u, P}(t), a^{\rm dis}_{u, P}(t)$] charge and discharge actions (in kW) for the \Acs{PEV} in  $u$ at  $t$

  \item[$z_u(t)$] overall power demand exceeding power limits (in kW) for  $u$ at  $t$

  \item[$y_{u, E}(t), y_{u, P}(t)$] binary variables, true if \Acs{ESS} and \Acs{PEV} 
    of  $u$ are charged in  $t$ and false otherwise 

  \item[$y_{u}^{low}(t), y_{u}^{high}(t), y_{u}^{\rm in}(t)$] binary variables, true if overall power $e_u(t)$ is greater than lower limit $P_u^{low}(t)$,  less than upper limit $P_u^{high}(t)$, inside lower and upper limits (resp.) in  $u$ at  $t$

\end{description}

\SectionInput[sec:intro]{Introduction}[intro]{intro.tex}

\SectionInput[sec:problem_form]{Problem Formulation and System Architecture}[problem_form]{problem_form.tex}

\SectionInput[sec:methodology]{Methodology}[methodology]{methodology.tex}

\SectionInput[sec:exps_setup]{Experimental Setup}[results]{exps_setup.tex}

\SectionInput[sec:exps_results]{Experimental Results}[results]{exps_res.tex}

\SectionInput[sec:conclusions]{Conclusions}[conclusions]{conclusions.tex}



\begin{IEEEbiography}[{\includegraphics[width=1in,height=1.25in,clip,keepaspectratio]{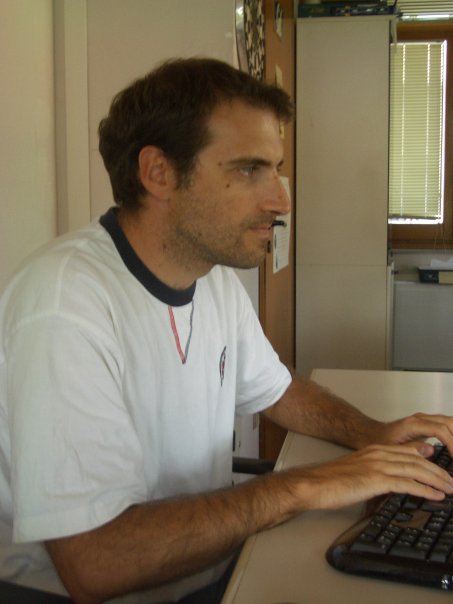}}]
{Igor Melatti} was born in Treviso, Italy. He received the Master Degree in Computer Science and the PhD in Computer Science and Applications from the University of L'Aquila, Italy in 2001 and 2005 respectively.
He has been a PostDoc at the University of Utah and at the Sapienza University of Rome. He is currently an Associate Professor at the Computer Science Department of Sapienza University of Rome.
His current research interests comprise: formal methods, automatic verification algorithms, model checking, software verification, cyber-physical systems, automatic synthesis of reactive programs from formal specification, systems biology, smart grids.
\end{IEEEbiography}

\begin{IEEEbiography}[{\includegraphics[width=1in,height=1.25in,clip,keepaspectratio]{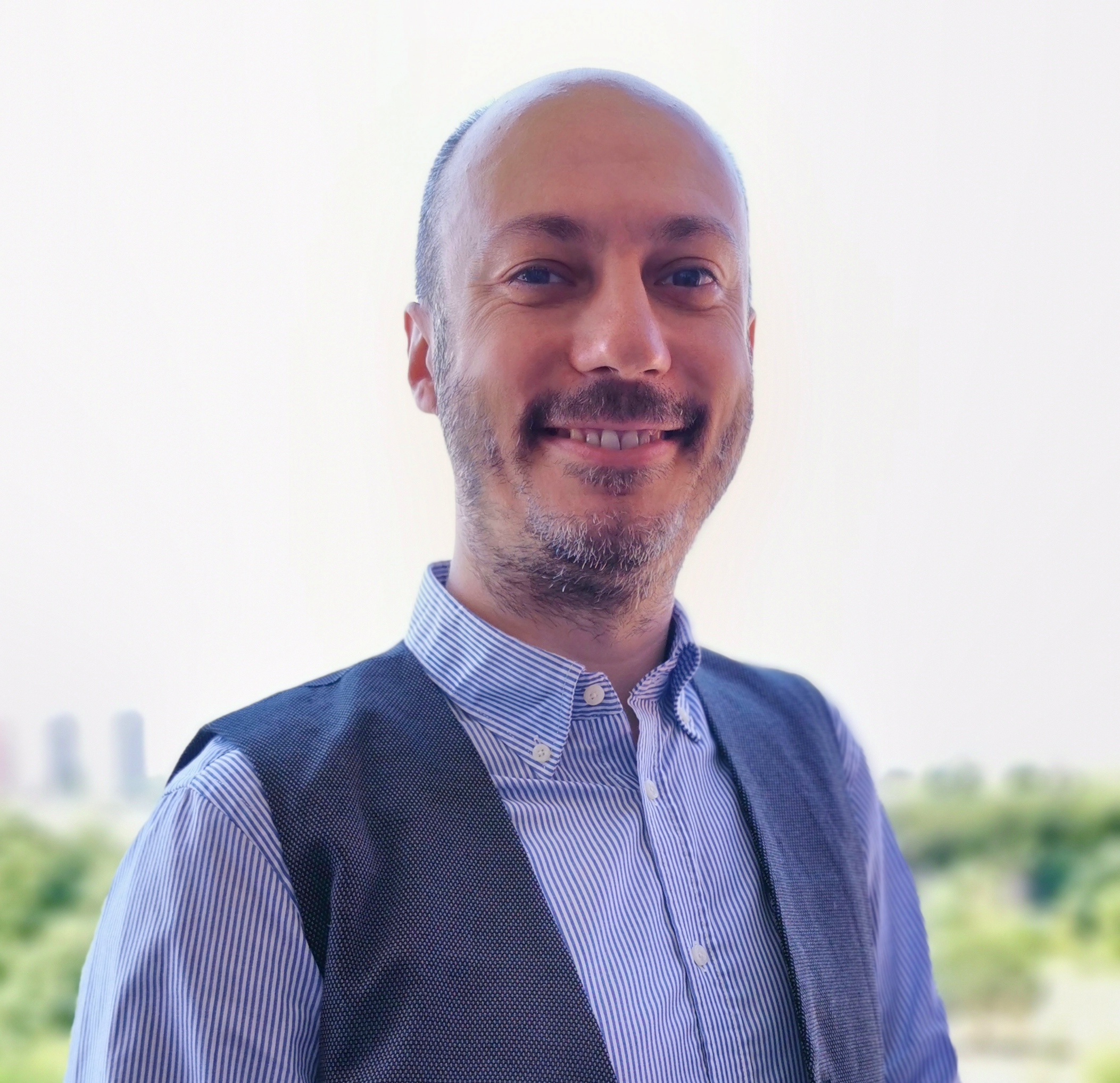}}]
{Federico Mari}
was born in Treviso, Italy. He received the Master Degree and the PhD in Computer Science from the Sapienza University of Rome, Italy in 2006 and 2010 respectively.
He is currently an Assistant Professor (tenure track) of Computer Science at the Department of Movement, Human and Health Sciences of the University of Rome Foro Italico (from 2019), where he serves as Rector's delegate for ICT.
His primary research interest is in formal methods and artificial intelligence applied to smart grids, health and sport sciences.
\end{IEEEbiography}

\begin{IEEEbiography}[{\includegraphics[width=1in,height=1.25in,clip,keepaspectratio]{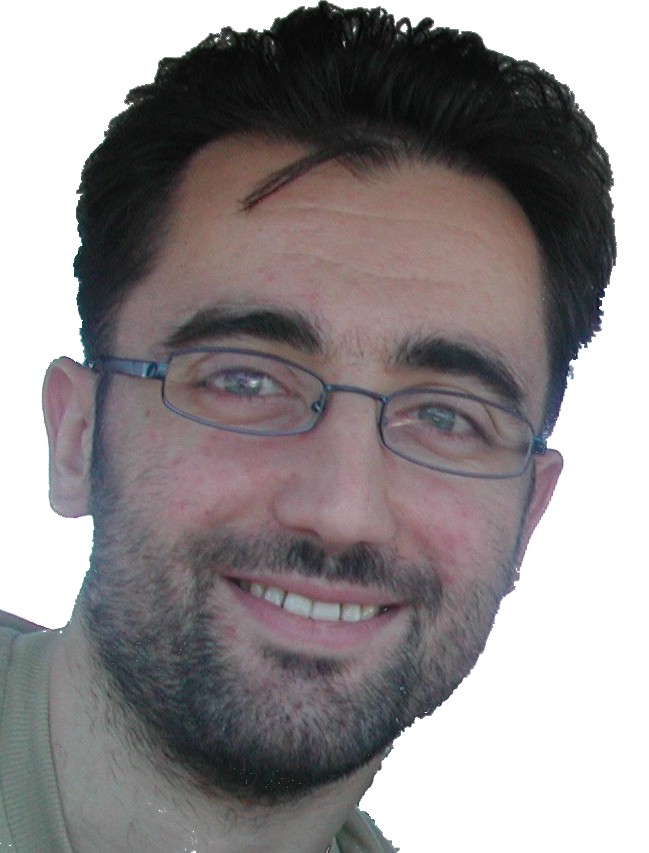}}]
{Toni Mancini}
has a Ph.D. in Computer Science Engineering and is Associate Professor at the Computer Science Department of Sapienza University of Rome, Rome, Italy. His research interests comprise: artificial intelligence, formal verification, cyber-physical systems, control software synthesis, systems biology, smart grids.
\end{IEEEbiography}

\begin{IEEEbiography}[{\includegraphics[width=1in,height=1.25in,clip,keepaspectratio]{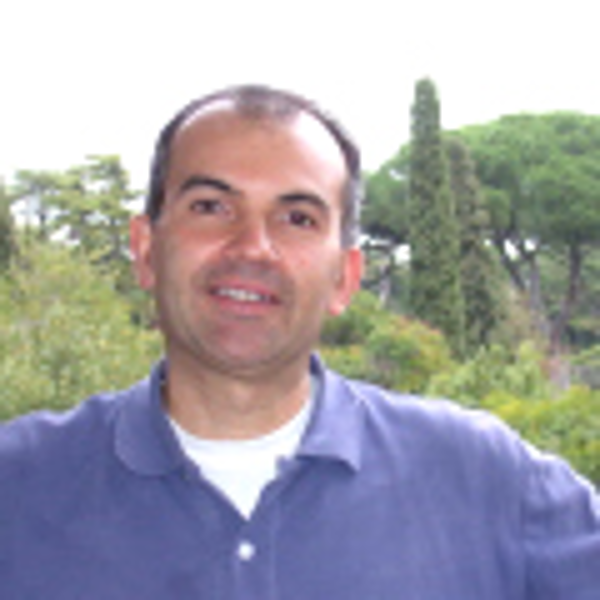}}]
{Enrico Tronci}
is a Full Professor at the Computer  Science  Department  of  Sapienza  University  of Rome (Italy). He received a master’s degree in Electrical Engineering from Sapienza University of Rome and a Ph.D.  degree  in  Applied  Mathematics  from  Carnegie Mellon  University.  His  research  interests  comprise: formal verification, model checking, system level formal verification, hybrid systems, embedded systems, cyber-physical systems, control software synthesis, smart grids, autonomous demand and response systems for smart grids, systems biology
\end{IEEEbiography}

\begin{IEEEbiography}[{\includegraphics[width=1in,height=1.25in,clip,keepaspectratio]{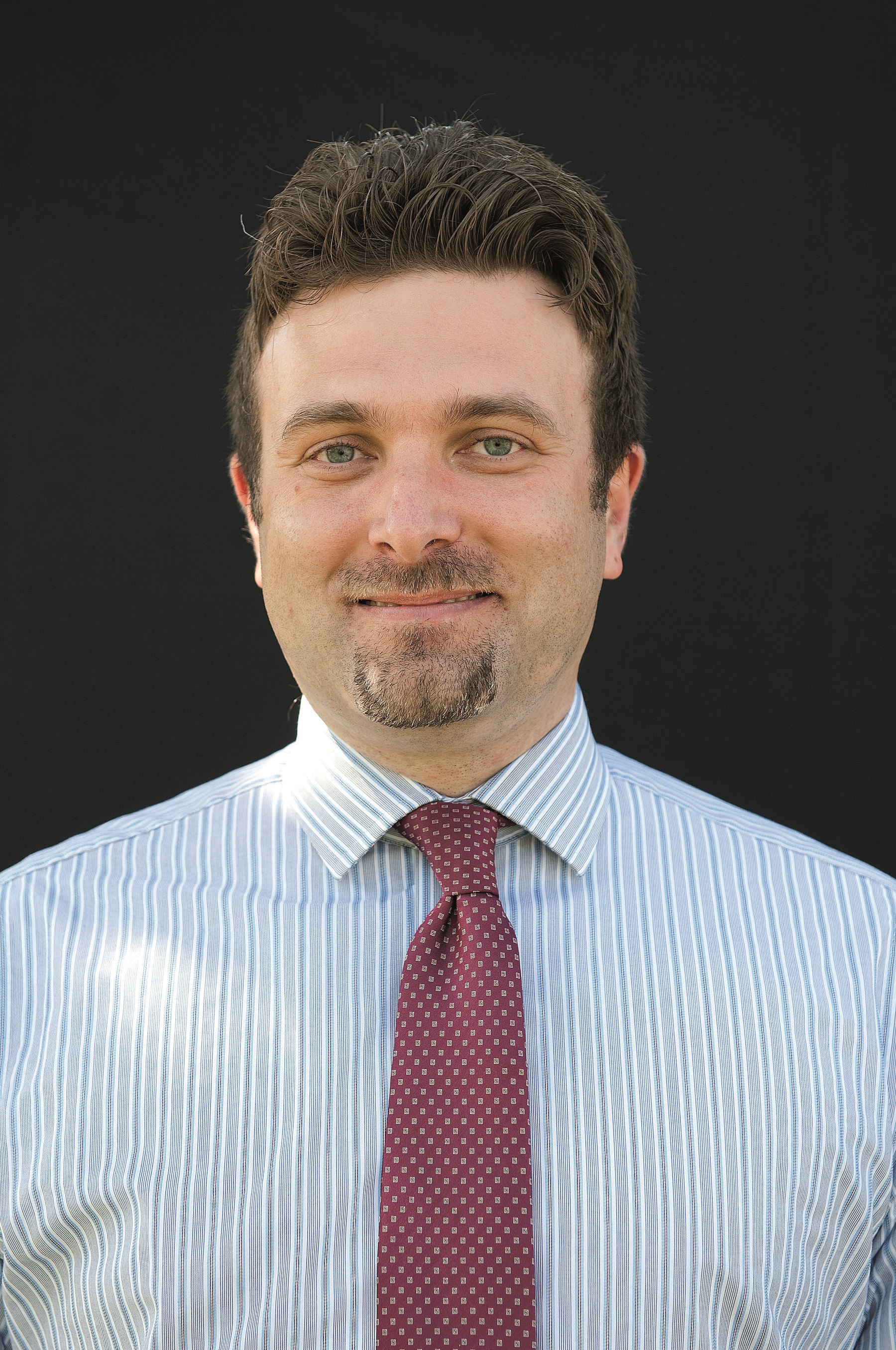}}]
{Milan Prodanovic} (M01) received the B.Sc. degree in electrical engineering from the University
of Belgrade, Serbia, in 1996 and the Ph.D. degree
from Imperial College, London, U.K., in 2004. From
1997 to 1999 he was engaged with GVS engineering
company, Serbia, developing UPS systems. From
1999 until 2010 he was a research associate in
Electrical and Electronic Engineering at Imperial
College, London, UK. Currently he is a Senior
Researcher and Head of the Electrical Systems Unit
at Institute IMDEA Energy, Madrid, Spain. His
research interests include design and control of power electronics interfaces
for distributed generation, micro-grids control and active management of
distribution networks.
\end{IEEEbiography}

\vspace{4cm}

\begin{mdframed}
\begin{center}
	\ArticleDisclaimer
\end{center}
\end{mdframed}

\end{document}